\documentclass[twocolumn,aps,showpacs,preprintnumbers,amsmath,amssymb,floatfix]{revtex4}

\usepackage{graphicx}

\begin{document}

\title{Localization and Interaction Effects in Strongly Underdoped
La$_{2-x}$Sr$_x$CuO$_4$}
\author{Marta Z. Cieplak,$^1$ A. Malinowski,$^1$ S. Guha,$^2$ and M. Berkowski$^1$}
\affiliation{
$^1$ Institute of Physics, Polish Academy of Sciences, 02 668 Warsaw,
Poland\\
$^2$ Department of Physics and Astronomy, Rutgers University, Piscataway, NJ
08855, USA}
\date{\today}

\begin{abstract}
{ The in-plane magnetoresistance (MR) in La$_{2-x}$Sr$_x$CuO$_4$ films with
$0.03 < x < 0.05$ has been studied in the temperature range 1.6 K to 100 K, and
in magnetic fields up to 14 T, parallel and perpendicular to the CuO$_2$ planes.
The behavior of the MR is consistent with a predominant influence of interaction
effects at high temperatures, switching gradually to a regime dominated by spin
scattering at low $T$. Weak localization effects are absent. A positive orbital
MR appears close to the boundary between the antiferromagnetic and the
spin-glass phase, suggesting the onset of Maki-Thompson superconducting
fluctuations deep inside the insulating phase. }
\end{abstract}

\maketitle

Doping of charge carriers drives the high-$T_c$ cuprates from the
insulating and antiferromangetic (AF) phase, to a spin-glass (SG),
and to a metallic and a superconducting (SC) phase. Understanding
the nature of this evolution is a fundamental but still
controversial problem. The properties of the underdoped regime are
very unusual, somehow related to the pseudogap opening observed by
angle resolved photoemission (ARPES)\cite{damasc}. Some
non-Fermi-liquid theories propose the existence of hidden order,
such as charge stripes, spin-charge separation, or orbital
currents \cite{sachdev}. On the other hand, a recent Fermi-liquid
model, with strong AF and SC fluctuations of Maki-Thompson (MT)
type, accounts well for transport anomalies of the pseudogapped
state \cite{kontani}.

A related puzzling, but less studied feature, is the conductance
in the SG region. It is metallic-like at high temperatures,
changing gradually into variable range hopping (VRH) at low $T$
\cite{kast,ando}. Recent ARPES experiment on the underdoped
La$_{2-x}$Sr$_x$CuO$_4$ (LSCO) reveals the emergence of sharp
nodal quasiparticle (QP) peak \cite{yoshida}. The peak appears
first in the SG at $x = 0.03$, and its spectral weight grows
linearly with the increase of $x$, suggesting that it is
responsible for the metallic conductance at high $T$. The question
arises what is the origin of the low-$T$ localization of the QP
states. Calculations show that a narrow QP band may arise in a
system with disordered charge inhomogeneities (stripes)
\cite{granath}, and there are suggestions that the charge stripes,
combined with the weak localization, may account for the
conductance behavior \cite{moshchalk}. However, before considering
these new approaches, one should study experimentally the low-$T$
localization and compare it to the standard localization and
interaction theories (LIT) for disordered systems \cite{leeram}.

In this study we probe, for the first time, the localization and
interaction effects in the SG regime of LSCO ($0.03 < x < 0.048$).
We use MR measurements in the longitudinal (LMR) and transverse
(TMR) configurations, with the magnetic field B parallel or
perpendicular to the $ab$ planes, respectively. In two-dimensional
(2D) systems the LMR usually probes spin-related effects, while
the TMR may, in addition, contain orbital contributions
\cite{leeram}. We find that the high-$T$ metallic-like conduction
displays features which are in qualitative agreement with the
standard LIT picture. The most important finding is the absence of
weak localization and the predominant influence of interactions.
The positive orbital MR suggests the onset of superconducting
fluctuations of the MT type deep inside the insulating phase,
confirming the importance of the MT effect suggested in Ref.
\cite{kontani}.

Our samples are $c$-axis aligned epitaxial films, about  6000 \AA\, thick, made
by pulsed laser deposition on substrates of LaSrAlO$_4$ \cite{cieplak1}. To
minimize the substrate-induced strain and the oxygen deficiency we select the
films with the lowest resistivity and smallest surface roughness from a large
body of specimens, as described in Refs. \cite{cieplak2,malin}. The films have
resistivities about 30 to 60\% larger than in the best bulk single crystals with
the same $x$ \cite{ando}. However, the superconductor-insulator (SI) transition
occurs in films at $x=0.05$, just as it does in single crystals. In addition, we
have found that the MR is insensitive to strain \cite{cieplak2,malin}, and the
MR in the films is the same as that in single crystals with the same $x$
\cite{balakir} so that we are led to believe that the data measured in these
films reflect the genuine behavior of the MR. The measurements were done in the
four-probe configuration, with magnetic fields up to 14 T, and temperatures down
to 1.6 K. The data were accumulated either by sweeping the field or by sweeping
the temperature, and the results were found to be consistent with one another.

\begin{figure}[ht]
\includegraphics[width=8cm]{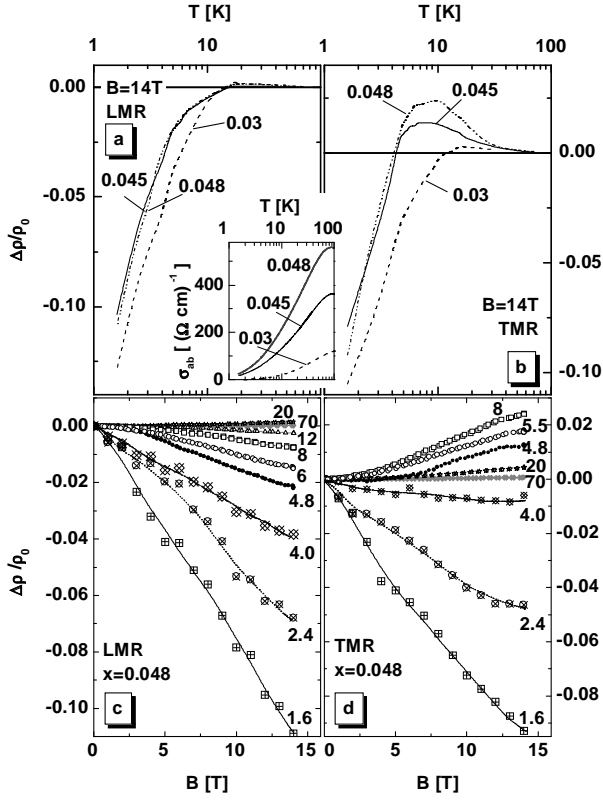}
\caption{The $T$-dependence of the LMR (a) and TMR (b) at the magnetic  field of
14 T for La$_{2-x}$Sr$_{x}$CuO$_4$ films with $x = 0.03, 0.045$, and 0.048. The
inset shows the $T$-dependence of the $\sigma_{ab}$ at zero magnetic field. (c)
LMR and (d) TMR as a function of $B$ for $x=0.048$ for several fixed
temperatures indicated in the figures. The data below 4.2 K were obtained from
temperature sweeps at fixed fields.\label{fig1} }
\end{figure}

\begin{figure}[ht]
\includegraphics*[scale=0.6]{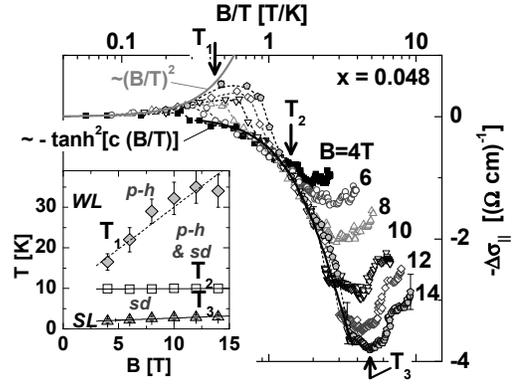}
\vspace*{0.0cm} \caption{ $-\Delta \sigma_{||}$ as a function of $\ln{(B/T)}$
for a film with $x = 0.048$. The data were taken during temperature sweeps  at
the constant fields indicated in the figure. The dotted lines are guides to the
eye. The arrows identify the characteristic temperatures $T_1$ to $T_3$ (defined
in the text) at 14 T. The solid lines, grey and black, show the common
dependencies followed by the data. The inset shows the field dependencies of
$T_1$, $T_2$ and $T_3$. The dashed line is fitted to $T_1 (B)$. \label{fig2} }
\end{figure}

The inset in Fig.~1 shows the $T$-dependence of the $ab$-plane conductivity,
$\sigma_{ab}$, at zero magnetic field, for the films with $x = 0.03, 0.045$, and
0.048, and $k_F l$ at 20K equal to 0.08, 0.35, and 0.57, respectively. Here $k_F
$ is the Fermi wave vector, $l$ is the mean-free path, and we use $k_F l =
hd{\sigma_{ab}}/e^2$ ($d$ is the distance between the CuO$_2$ planes). This
formula underestimates $k_F l$ for nodal QP, ARPES shows that $k_F l$ around
node is much larger \cite{yoshida}. The main figures 1(a) and 1(b) show the
$T$-dependence of the MR, defined as $\Delta \rho /{\rho_0} = (\rho (B) - \rho_0
)/ \rho_0 $, where $\rho_0$ is the resistivity at zero field, measured in the
LMR and TMR configurations. Below about 3K $\sigma_{ab} (T)$ approaches the Mott
VRH law and both the LMR and TMR are large and negative. The appearance of the
negative LMR coincides approximately with the development of the magnetic
quasi-elastic neutron scattering intensity in the SG phase \cite{wakimoto}. The
magnitude of the negative MR is similar in two films with larger strontium
contents, and it is substantially larger for $x = 0.03$. This suggests that the
negative MR is a spin-related effect, which becomes stronger as the AF
correlations grow with the decrease of $x$. At higher $T$ there is a gradual
transition to a weaker $T$-dependence of the conductivity, $\sigma_{ab} \propto
\ln{T}$. The slope $S_0 = d(\sigma_{ab})/d(\ln{T})$ increases by a factor of 3.5
when $x$ grows from 0.03 to 0.048. This is accompanied by the gradual appearance
of a positive contribution to both LMR and TMR. The TMR is always larger than
the LMR, and the difference increases as $x$ grows. This indicates that there is
a substantial orbital contribution to the MR (OMR) related to the presence of
delocalized carriers at high $T$. The existence of the positive OMR is also
evident from the dependencies of the LMR and the TMR on the magnetic field,
which are shown in Figs.~1(c) and 1(d) for $x=0.048$ for several fixed
temperatures. By analogy with conventional disordered metals we call the
high-$T$ range, the "weakly localized regime" (WL), and the low-$T$ range, the
"strongly localized regime" (SL). In the following we focus primarily on the WL
regime.

Fig.~2 shows the LMR data for $x=0.048$, replotted as a magnetic-field  induced
change of the conductivity, $-\Delta \sigma_{||} = {\Delta \rho}_{||}
/{\rho_0}^2$, versus $B/T$. Qualitatively similar behavior is observed for
different $x$. We define three characteristic temperatures, $T_1$ - $T_3$, which
are shown in Fig.~2 for the data taken in a field of 14 T. At $T > T_1$, which
corresponds to $B/T << 1$, all data for different fields follow one common
curve, $-\Delta \sigma_{||} = S_L \;(B/T)^2$ (solid grey line). The parameter
$S_L$ increases by a factor of about 5 as $x$ grows from 0.03 to 0.048, an
increase by about the same order of magnitude as the increase of $S_0$
\cite{param}. At $T_1$ the positive contribution deviates from the quadratic
dependence and saturates. This is followed by the appearance of a negative
contribution, until at the temperature $T_2$ the data join a second common
curve, best described by a simple functional form $\Delta \sigma_{||} \sim
{\tanh}^2 [c \, (B/T)]$ (solid black line). At the lowest $T$ the negative LMR
deviates from this curve and a new anomaly appears at $T_3$.

The inset to Fig.~2 shows the field dependence of $T_1$, $T_2$, and $T_3$. $T_2$
and $T_3$, which are related to the negative contribution of the LMR, are
field-independent. Apparently the spin-related scattering is influenced mainly
by the temperature, not by the external magnetic field. This suggests the
presence of strong internal fields and is consistent with SG ordering driven by
the temperature-induced localization of carriers. It is most likely that the
low-$T$ LMR results from the suppression of spin-disorder (sd) scattering by the
magnetic field, that the simple functional form reflects the dependence of the
local magnetization on $B/T$, and that the reduction of the scattering rate
below the SG freezing temperature leads to the anomaly around $T_3$. More
discussion of this behavior will be presented elsewhere.

\begin{figure}[ht]
\includegraphics[width=7.8cm]{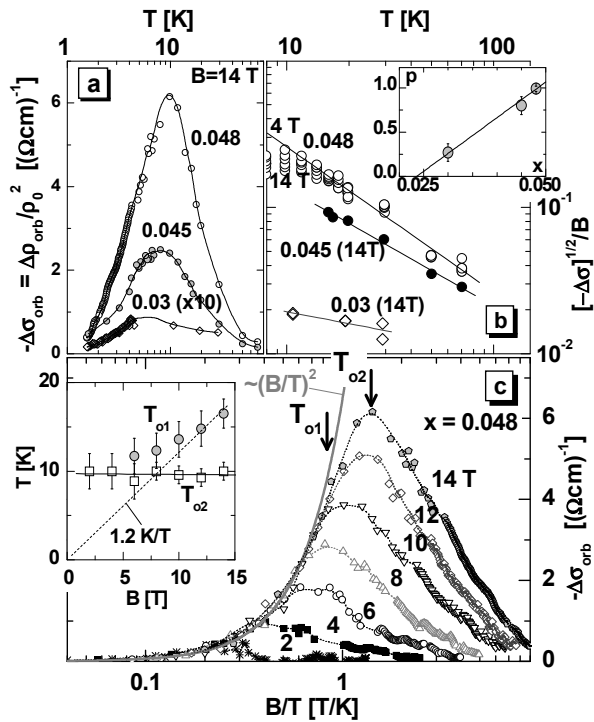}
\caption{ (a) $-{\Delta \sigma}_{orb}$ as a function of $T$ for films with $x =
0.03$, 0.045, and 0.048 at a magnetic field of 14 T. The  data for $x = 0.03$
are multiplied by 10. (b) $S_{orb}^{1/2}=[-{\Delta \sigma}_{orb}]^{1/2}/B$ as a
function of $T$ on a double logarithmic scale. The data for $x = 0.048$ are
shown for several magnetic fields between 4 T and 14 T, and the data for $x =
0.03$ and $ x = 0.045$ are for a field of 14 T. The solid lines are fitted to
the data at 14 T and at $T > 16$ K. The inset in (b) shows the exponent $p$ as a
function of $x$. (c) $-{\Delta \sigma}_{orb}$ as a function of $\ln{B/T}$ for $x
= 0.048$. $T_{o1}$ and $T_{o2}$ are crossover temperatures shown for a field of
14 T. The inset in (c) shows the field dependencies of  $T_{o1}$ and $T_{o2}$.
The dashed line is the slope of the dependence $T_{o1} (B)$ for high magnetic
fields. \label{fig3} }
\end{figure}

On the other hand $T_1$ increases with $B$. We associate the positive
contribution in the WL regime with the effect of Zeeman splitting in the
particle-hole (p-h) interaction channel. According to LIT \cite{leeram}, in a 2D
system this effect should be described by the dependence $\Delta \sigma \sim -
g_2(h)$, where $h = g{\mu_B}B/k_B T$, and $g_2(h)$ is a function which has
limiting behaviors $g_2 = 0.084 h^2$ for $h << 1$, and $g_2 = \ln{(h/1.3)}$ for
$h >> 1$. Therefore, we expect the deviations from the $(B/T)^2$ curve to occur
when $g{\mu_B}B \sim k_B T$, i.e. when $T/B = g{\mu_B}/k_B = g \times 0.67$ K/T.
The straight line fitted to the dependence of $T_1$ on $B$ has a slope $T_1 /B$
of about $(2.0 \pm 0.4)$ K/T, giving an effective $g$-factor of about $(3 \pm
0.6)$, enhanced in comparison with the free-electron value. The fitted line has
a finite intercept, which is not predicted by the theory. The deviations from
theory are probably the results of the increasing influence of sd scattering as
the temperature $T_2$ is approached from above, although the enhancement of $g$
may be also a genuine effect caused by strong internal fields in the SG phase.

Next we consider the TMR. If the spin-orbit coupling is small, one expects that
the LMR is isotropic. Assuming tentatively that this is the case, we extract the
OMR by subtracting the LMR from the TMR. Fig.~3a shows that ${-\Delta
\sigma}_{orb}$ is positive, and has the form of a pronounced maximum, which
rapidly increases with an increase of $x$. This OMR of the insulating samples
evolves smoothly into a large OMR for SC films with $x > 0.06$ which we have
studied previously \cite{balakir}. In SC films the OMR diverges above the SC
transition temperature. A recent Fermi liquid theory explains the OMR
divergence, together with other anomalous transport properties below the
pseudogap opening, by the interplay of strong AF and SC fluctuations of the MT
type \cite{kontani}. The smooth evolution across the SI transition suggests that
the origin of the positive OMR may be similar below and above the SI transition,
possibly related to MT fluctuations.

Fig.~3c shows ${-\Delta \sigma}_{orb}$ as a function of $B/T$ for $x = 0.048$.
At the highest temperatures, which correspond to small $B/T$, the data follow a
$(B/T)^2$-dependence. Below the temperature $T_{o1}$ a crossover occurs to a
weaker dependence, and at a still lower temperature, $T_{o2}$, the OMR decreases
rapidly. The field dependencies of $T_{o1}$ and $T_{o2}$, shown in the inset,
reveal that $T_{o2}$ is constant. In fact, comparison with Fig.~2 tells us that
$T_{o2}$ closely matches the temperature $T_2$, below which sd scattering
dominates the LMR. We conclude that the low-$T$ suppression of the OMR is
associated with strong sd scattering. On the other hand $T_{o1}$ increases with
increasing field.

When $x$ decreases, ${-\Delta \sigma}_{orb}$ is still proportional
to $B^2$, but the $T$-dependence changes. We derive the relation
for each $x$, $-{\Delta \sigma}_{orb} = S_{orb}(T)\; B^2 $, where
$S_{orb} (T)$ is the coefficient which depends on temperature as a
power-law, $S_{orb} \sim T^{-2p}$, with the exponent $p$ which
changes with $x$. In Fig.~3b we plot $S_{orb}^{1/2}=[-{\Delta
\sigma}_{orb}]^{1/2}/B$ as a function of $T$ on a double
logarithmic scale. Note first that $S_{orb}$ decreases by two
orders of magnitude when $x$ drops from 0.048 to 0.03. This may be
compared with the change of $S_0$ or $S_L$, both of which decrease
by a much smaller factor. This shows that the suppression of the
OMR is not simply related to the decrease of concentration of
delocalized carriers, but involves a process which is more
strongly dependent on $x$. The slope of the linear dependence
gives the exponent $p$, which is shown as a function of $x$ in the
inset to Fig.~3b. It reduces approximately linearly with a
decrease of $x$, from $0.99 \pm 0.03$ for $x = 0.048$, to $0.27
\pm 0.1$ for $x = 0.03$.  Extrapolating the linear dependence $p
(x)$ to $p = 0$ we get the value of $x$ above which the OMR is
observed, $x_c = 0.023$. This is close to $x = 0.02$, which is the
boundary between the AF phase and the SG phase. Apparently the OMR
is detectable as soon as the concentration of carriers is large
enough to frustrate the long-range AF order.

According to LIT \cite{leeram} a positive OMR may be caused by weak localization
in the presence of strong spin-orbit scattering. However, a change of sign of
the OMR for high magnetic fields is then expected, and we see no trace of that.
Another possible source of the positive OMR is the particle-particle (p-p)
interaction channel. A particular type of p-p scattering is analogous to the MT
process of electron interactions with SC fluctuations \cite{larkin}. The
magnitude of this effect is very small for normal metals with repulsive
interactions, but may be large in systems with attractive interactions. The
functional form of the OMR is the same as that caused by weak localization, with
two distinct dependencies on the magnetic field, $ -\Delta \sigma \sim
(B/B_{in})^2 $ for $B << B_{in}$ and $ -\Delta \sigma \sim \ln(B/B_{in})$ for $
B >> B_{in} $. Here $B_{in} = \hbar /4eD\tau_{in}$, where $D$ is the diffusion
coefficient, and $\tau_{in}$ is the inelastic scattering time, usually given by
$\tau_{in} \sim T^{-p}$, with an exponent $p$ which depends on the scattering
mechanism.

These predictions would be consistent with our experiment, if we
interpret the $T_{o1} (B)$-line as a crossover between distinct
dependencies on $B/B_{in}$. To avoid the influence of sd
scattering we fit the straight line to the $T_{o1} (B)$ data at
highest magnetic fields. We get $T/B_{in} = 1.2$ K/T, and from
this we obtain $D\tau_{in} = 198 T^{-1}$ nm$^2$. The inelastic
scattering length, $L_{in} = (D \tau_{in})^{1/2}$, is about 30
{\AA} at 20 K. The decrease of $x$ to 0.03 reduces this distance
to about 10 {\AA}. This may be compared to the mean-free path
estimated from ARPES experiment, 16 {\AA} at 20 K for $x=0.03$
\cite{yoshida}. The agreement between these two experiments is
very good. Using the diffusion coefficient (from conductivity) for
$x=0.048$ film, $D \approx 0.39$ cm$^2$/s at 20 K we get the
inelastic scattering time $\tau_{in} \approx 2.6 \times 10^{-13}$
s (this should be treated as an upper bound since $D$ from
conductivity is underestimated). The magnitude of $\tau_{in}$ is
at least two orders of magnitude smaller than in conventional
disordered metals, revealing much stronger interactions. On the
other hand, the exponent $p = 1$ for $x=0.048$ is consistent with
the expectations for electron-electron interactions in a
disordered 2D metal \cite{leeram}. The reduction of $p$ with the
decrease of $x$ is anomalous, pointing to an unusual strong
scattering weakly dependent on $T$.

The previous MR experiments on metallic (but nonsuperconducting) high-$T_c$
systems found a power-law $T$-dependence of $\tau_{in}$ with small $p$, close to
our results for $x = 0.03$, but the OMR was negative, caused by weak
localization \cite{jing,hagen}. It is tempting to speculate that the small value
of $p$ may be a generic feature of a narrow nodal QP band, which becomes evident
in the absence of the MT process. However, the differences between the previous
cases and the LSCO system may be more profound, including, for example, the
differences in the electronic structure and the opening of the pseudogap, the
magnitude of spin-related scattering, etc. More experiments on the SG samples
for different cuprates are needed to search for similar OMR behavior in other
systems.

We now consider what are the implications of these results for the understanding
of the SI transition in cuprates. First, the behavior of the positive OMR
suggests that it originates in the MT contribution, and this is in a very good
agreement with the Fermi-liquid description of the pseudogap state proposed in
Ref.\cite{kontani}. However, it is not yet clear how far into the insulating SG
phase this description may be extended. In addition, our discussion is based on
the qualitative comparison with the standard LIT, which may not exactly hold for
the narrow QP band. Secondly, our results indicate that the model of conducting
stripes combined with weak localization effect is not sufficient to describe the
conductance in strongly underdoped LSCO. On the other hand, it should be
stressed that the appearance of the positive OMR deep in the insulating phase
does not rule out the theories based on the charge inhomogeneity picture. One
can imagine that the MT fluctuations may be enhanced in the regions with high
concentration of carriers. More elaborate models based on a stripe picture are
needed, incorporating the presence of MT fluctuations.

In conclusion, we have revealed the origin of the effective localization of the
QP states at low temperatures in the strongly underdoped LSCO. The MR experiment
shows that the weak localization effect is absent. Instead, these are the
interaction effects which play the decisive role. The LIT picture \cite{leeram}
gives a good qualitative description of the MR in the WL regime, reproducing the
dependence on the field and temperature. It suggests the onset of the MT
fluctuations at the AF-SG phase boundary.

We appreciate discussions with T. Dietl and P. Lindenfeld. This work was
supported  by the Polish Committee for Scientific Research, KBN, grant 2 P03B
044 23.


\begin{thebibliography}{99}
\bibitem{param}
In conventional disordered systems the parameters $S_L$ and $S_0$ and the LIT
equations [9] could be used to estimate the relative contribution of the
localization and interaction effects. However, numerical coefficients in these
equations will be probably modified in case of narrow nodal QP band and it is
not immediately obvious to us how to modify them.
\end{thebibliography}
\end{document}